\documentclass[10pt, conference]{IEEEtran}
\IEEEoverridecommandlockouts

\usepackage{multicol}
\usepackage{amsmath, amssymb, bm, cite, epsfig, psfrag, mathtools}
\usepackage{epstopdf}
\usepackage{changepage}
\usepackage{graphicx}
\usepackage{fancyhdr}
\usepackage{bbm}
\usepackage{algorithm,algpseudocode}
\usepackage{array}
\usepackage[margin=10pt,font=small]{caption}
\usepackage{bbm}
\usepackage{multirow}
\usepackage[usenames,dvipsnames]{xcolor}

\usepackage{subfigure}
\usepackage{etoolbox}
\usepackage{pbox}
\usepackage[width=0.48\textwidth]{caption}
\usepackage{colortbl}
\usepackage{bm}
\usepackage{caption}
\usepackage{amsmath}
\usepackage{color}

\usepackage{enumitem}
\usepackage{datetime}
\usepackage{amsthm}
\usepackage{dsfont}
\usepackage{mathtools}
\usepackage[T1]{fontenc}

\newtoggle{conference}
\togglefalse{conference} 
\usepackage[letterpaper,left=0.75in,right=0.75in,top=0.75in,bottom=0.75in,footskip=.25in]{geometry}
\graphicspath{{figures/}}

\newcommand{\eqdef}{\stackrel{\Delta}{=}}

\def\beq{\begin{equation}}
\def\eeq{\end{equation}}
\def\beqa{\begin{eqnarray}}
\def\eeqa{\end{eqnarray}}
\def\beqan{\begin{eqnarray*}}
\def\eeqan{\end{eqnarray*}}

\def\Pm{{\mathbb{P}}}

\def\argmax{\mathop{\mathrm{arg\,max}}}

\newtheorem{theorem}{Theorem}

\def\dsf{\sf {d}}

\def\tm1{t\! - \! 1}
\def\tp1{t\! + \! 1}

\def\dbf{\mathbf{d}}

\def\fbf{\mathbf{f}}

\def\pbf{\mathbf{p}}

\def\qbf{\mathbf{q}}

\def\Qbf{\mathbf{Q}}
\def\Mbf{\mathbf{M}}

\def\Uc{\mathcal{U}}

\def\E{{\mathbb E}}

\begin{document}

\bibliographystyle{IEEEtran}

\title{\vspace{0.25in}Distortion-Memory Tradeoffs in Cache-Aided Wireless Video Delivery}

\author{
P. Hassanzadeh, E. Erkip, J. Llorca, A. Tulino
\thanks{P. Hassanzadeh  and  E. Erkip are with the ECE Department of New York University, Brooklyn, NY. Email: \{ph990, elza\}@nyu.edu}
\thanks{J. Llorca  and A. Tulino are with Bell Labs, Alcatel-Lucent, Holmdel, NJ, USA. Email:  \{jaime.llorca, a.tulino\}@alcatel-lucent.com}
\thanks{A. Tulino is with the DIETI, University of Naples Federico II, Italy. Email:  \{antoniamaria.tulino\}@unina.it}
}

\maketitle


\begin{abstract}
Mobile network operators are considering caching as one of the strategies to keep up with the increasing demand for high-definition wireless video streaming. By prefetching popular content into memory at wireless access points or end user devices, requests can be served locally, relieving strain on expensive backhaul. In addition, using network coding allows the simultaneous serving of distinct cache misses via common coded multicast transmissions,  
resulting in significantly larger load reductions compared to those achieved with conventional delivery schemes. However, prior work does not exploit the properties of video and simply treats content as fixed-size files that users would like to fully download. Our work is motivated by the fact that video can be coded in a scalable fashion and that the decoded video quality depends on the number of layers a user is able to receive. Using a Gaussian source model, caching and coded delivery methods 
are designed to minimize the squared error distortion at end user devices.
Our work is general enough to consider heterogeneous cache sizes and video popularity distributions.
\end{abstract}


\section{Introduction}~\label{sec: Introduction}
With the recent explosive growth in cellular video traffic, wireless operators are heavily investing in making infrastructural improvements such as increasing base station density and offloading traffic to Wi-Fi.
Caching is a technique to reduce traffic load by exploiting the high degree of asynchronous content reuse and the fact that storage is cheap and ubiquitous in today's wireless devices~\cite{websiteintel,golrezaei2011wireless,molisch2014caching}. During off-peak periods when network resources are abundant, popular content can be stored at the wireless edge (e.g., access points or end user devices), so that peak hour demands can be met with reduced access latencies and bandwidth requirements.

The simplest form of caching is to store the most popular video files at every edge cache~\cite{LRUcaching}. Requests for popular cached files can then be served locally, while cache misses need to be served by the base station, achieving what is referred to as a 
local caching gain. However,  
replicating the same content on many devices can result in an inefficient use of the aggregate cache capacity~\cite{LRUnotGood}.
In fact, recent studies~\cite{CACHE, maddah2013decentralized, ji2014average, ji2015random, Pr:JiTulLloCai14} have shown that making users store different portions of the video files creates coded multicast opportunities that enable a global caching gain.
In \cite{maddah2013decentralized}, it is shown that uniform random caching, in which users cache portions of every file uniformly at random, in combination with linear index coding, achieves a worst-case rate that is within a constant factor of an information theoretic lower bound; and hence, is order-optimal.
The case of random demands 
according to a Zipf popularity distribution is analyzed in \cite{ji2014average, ji2015random}. The authors characterize the optimal average rate as a function of all system parameters and provide an order-optimal caching and coded multicast scheme 
designed to balance the gains from local cache hits and coded multicast opportunities.

While existing work on wireless caching is motivated by video applications, specific properties of video are not exploited in the caching and delivery phases.
In scalable video coding (SVC)~\cite{svc}, video files are encoded into layers such that the base layer contains the lowest quality level and additional enhancement layers allow successive improvement of the video streaming quality.

In this work, we analyze the use of caching as a method to enhance video quality at users' streaming devices.
We consider a scenario in which users store videos at different encoding rates (e.g., video layers in SVC).
Upon video streaming requests, depending on the available network resources, users receive additional layers that successively refine the video playback quality. To formulate the problem mathematically, we assume that the library consists of Gaussian sources with different variances. We allow for users to have different cache sizes and probability distributions for accessing these sources.
The goal is to design caching and delivery schemes that, for a given broadcast rate, minimize the average distortion experienced at user devices.
We first show that under unicast delivery, the optimal caching policy admits a reverse water-filling type solution which can be implemented locally and independently across users, without the need of global coordination. Each video streaming request just needs to specify the maximum quality level available at the corresponding local cache, such that the sender (e.g., base station) can effectively compute the optimal delivery rates allocated to each user.
We then show how using coded multicast offers notable performance improvements in terms of average video distortion.
In this case, the optimal policy requires joint optimization of the rates at which videos are cached by each user. After users place their requests, the sender, with knowledge of the users' cache contents, computes a common multicast codeword that simultaneously delivers additional enhancement layers to each user.
Our simulation results confirm the significant gains achievable via coordinated caching and coded multicast, with more than $10\times$ reduction in average distortion observed in wireless caching networks with $20$ user caches and $100$ videos.

The remainder of this paper is organized as follows. The problem setting is introduced in Sec. \ref{sec: ProblemSetting}. The use of local caching and unicast transmission is analyzed in Sec. \ref{sec: Unicast}.  Sec. \ref{sec: Multicast} describes the proposed achievable scheme based on cooperative caching and coded multicast transmissions. Finally, numerical results that illustrate the achievable distortion-memory tradeoffs are presented in Sec. \ref{sec:Simulations}.

\section{Problem Setting}~\label{sec: ProblemSetting}
\begin{figure}[t]
  \centering
  \includegraphics[width=2.0in]{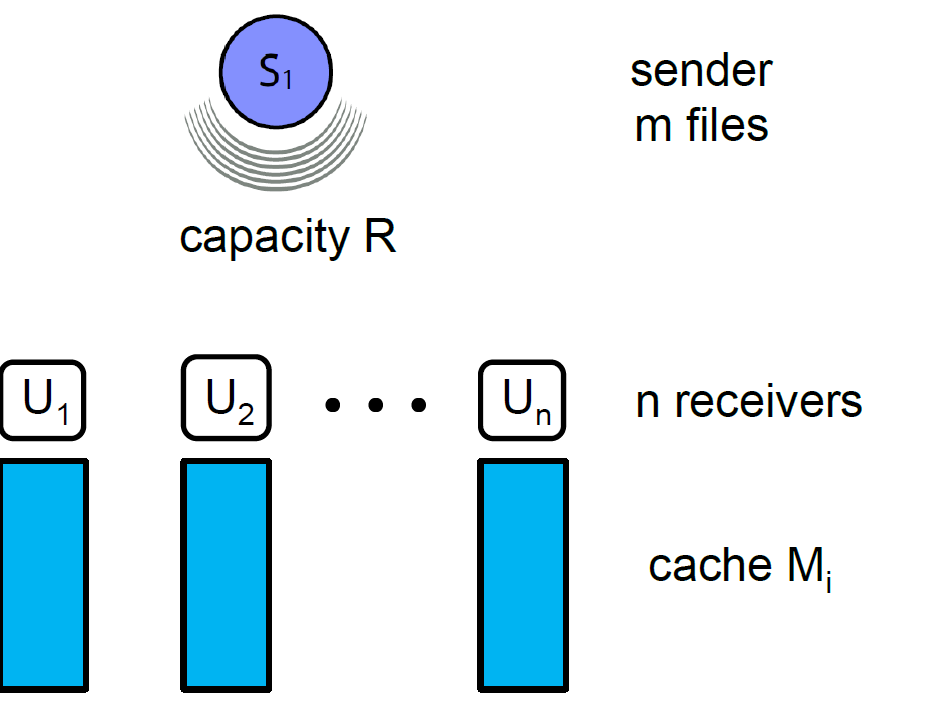}
  \caption{System Model. Caching is used for improving video playback qualities.}
  \label{fig: System}
\end{figure}

Consider the system in Fig.~\ref{fig: System} where one sender (e.g., base station) is connected through an error-free shared link to $n$ receivers (e.g., access points or user devices) with rate (capacity) $R$ bits/source-sample.
The sender has access to a content library, $\mathcal{F}=\{1,\ldots,m\}$, containing $m$ video files (sources) each composed of $F$ source samples.
Receiver $i\in\{1,\ldots,n\}$ has a cache of size $M_{i}$ bits/source-sample, or equivalently, $M_{i}F$ bits.

Receivers place requests for videos in the library according to a \textit{demand distribution} $\Qbf = [q_{i,j}],\; i=1, \cdots, n,\; j=1, \cdots, m$, assumed to be known at the sender,
where $q_{i,j} \in [0,1]$ and $\sum_{j=1}^m q_{i,j} = 1, \; \forall i\in[n]$\footnote{Throughout the rest of this paper, $[n]$ denotes the discrete set of integers from $1$ to $n$ i.e. $[n] \eqdef \{1,...,n\}$}. The demand distribution is defined such that receiver $i$ requests video file $j$ with probability $q_{i,j}$.
We use $\dsf_{i}$ to denote the random request at receiver $i$, with $d_{i}\in \mathcal{F}$ being a realization of $\dsf_{i}$.

We consider a video streaming application, in which each file $f\in\mathcal F$ represents a video segment, compressed using SVC \cite{svc}.
In SVC, the base layer contains the lowest level of detail spatially, temporally, and from a quality perspective. Additional layers, named enhancement layers, can improve the quality of the stream.
 Note that an enhancement layer is useless, unless the receiver has access to the base layer and all preceding enhancement layers. The decoded video quality depends on the total number of layers received in sequence.\\

The video delivery system operates in two phases, a caching (or placement) phase followed by a transmission phase:

\begin{itemize}
\item{\textbf{Caching Phase:}}
The caching phase occurs during a period of low network traffic. In this phase, all receivers have access to the entire library for filling their caches. In the following, without loss of generality, and in line with~\cite{ji2014average, ji2015random, Pr:JiTulLloCai14}, we assume that the caching phase is described by a set of $n$ vectors,  $\pbf_i=[p_{i,1},\ldots, p_{i,m}]$, $i\in[n]$, referred to as the caching distributions, with $\sum_{j=1}^m p_{i,j} = 1, \; \forall i\in[n]$.
Element $p_{i,j}$ represents the cache portion assigned to video $j$ at receiver $i$, and $M_{i,j}\eqdef p_{i,j} M_i$ is the number of bits/source-sample  of video $j$  cached at receiver $i$. We refer to the $m$-dimensional vector, $\Mbf_i = [M_{i,1},\ldots, M_{i,m}]$, as the \textit{cache placement} of user $i$.

Designing the caching phase consists of designing caching distributions, $\pbf_i \; \forall i \in[n]$, and the corresponding cache content. This can be done locally by the receivers, based on their local information, or globally (in a cooperative manner either directly by the sender, or by the receiver itself) based on information from the overall network. As in~\cite{CACHE, maddah2013decentralized, ji2014average, ji2015random, Pr:JiTulLloCai14}, we assume that library files and their popularity change at a much slower time-scale than the video delivery time-scale, and neglect the resource requirements associated with the cache-update process. Hence, the caching phase is sometimes referred to as the placement or prefetching phase.

\item{\textbf{Transmission Phase:}}
The transmission phase takes place after completion of the caching phase. During this phase, only the sender has access to the library. The network is repeatedly  used in a time slotted fashion.
At the beginning of each time slot, each receiver requests a video file $d_i \in \mathcal{F}$.
Having been informed of the demand realization $\mathbf{d}\eqdef [d_1, d_2, \ldots, d_n]$, the sender decides on the playback qualities of the requested videos. The sender computes the rate, $R_{i,\dbf}$ (bits/source-sample), that will be transmitted to receiver $i \in [n]$, based on the demand, cached content, and the channel capacity constraint. We refer to $R_{i,\dbf}$ as the per-user rate for demand realization $\mathbf{d}$.
 The sender encodes the chosen video layers into a codeword, ${\bf X}_\mathbf{d}$, which is then transmitted to the receivers over the shared link.
Receiver $i$ decodes its video of interest, $d_{i}$, (at the corresponding quality level) using the received codeword, ${\bf X}_\mathbf{d}$, and its cache content. Receiver's requested video playback quality depends on both $M_{i,d_i}$ and $R_{i,\dbf}$.



 \end{itemize}

For ease of exposure and analytical tractability, we assume that video file $j$ consists of $F$ i.i.d. Gaussian samples with variance $\sigma_{j}^{2}$ and distortion-rate function $D_{j}(r) = \sigma_{j}^{2} 2^{-2r}$\cite{cover}.
In this setting, based on the fact that Gaussian sources with squared error distortion are successively refinable with a multiple-stage description that is optimal at each stage \cite{successive}, scalable coding does not have any coding overhead.

The goal is to design the caching and transmission schemes that minimize the expected distortion (over the demand distribution), defined as
\begin{eqnarray}
\E(D) = \sum_{ \dbf \in \mathcal D } \Pi_{\bf d} \bigg(\frac{1}{n}  \sum_{i=1}^{n}\sigma_{{d}_{i}}^{2}2^{-2(M_{i,d_i}+ R_{i, \dbf})}\bigg).
\label{averDist}
\end{eqnarray}
In \eqref{averDist}, $\mathcal D$ is the set of possible demands,  $\dbf \in \mathcal D $  represents the specific demand realization, and $\Pi_{\bf d}$ is the probability of demand $\dbf$.
We  assume that receivers request video files independently; hence,  $\Pi_{\dbf} { \eqdef }\prod_{i=1}^n q_{i,d_{i}}$.


Note that, in this paper, the goal is to minimize the expected distortion when receivers are connected to the sender through a shared link of finite capacity, $R$ bits/source-sample. This is in contrast to prior work, {\cite{CACHE,maddah2013decentralized,ji2014average, ji2015random, Pr:JiTulLloCai14,ji2015Efficient,ji2015Preserving}, in which the goal is to minimize the total rate transmitted over the shared link, in order to recover all requested fixed-size (and hence fixed-quality) video files.


In the following, we focus on two scenarios for designing the caching and transmission schemes, that differ in performance, computational complexity and required coding overhead. Specifically, in the first scenario, in order to limit the computational complexity and reduce the communication overhead, we assume that the sender compresses the requested videos independently for each receiver, merely based on their local cached content.
We refer to such an encoding strategy 
(caching \& transmission scheme), as the \textit{Local Caching-aided Unicast (LC-U)} scheme.
On the other hand, in the second scenario, we assume that the sender compresses the requested videos jointly across all receivers based on global network knowledge (cache contents and demand distributions).
We refer to this second encoding strategy as the \textit{Cooperative  Caching-aided  Coded Multicast (CC-CM)} scheme.

Optimization of the caching and transmission phases for these two schemes are conducted differently and require different levels of complexity.
Note that CC-CM uses global network knowledge to construct codes that fully exploit the multicast nature of a wireless system, while LC-U eliminates additional coding complexity, but the total wireless resource (rate) must be orthogonally divided among receivers.



\section{Local Caching-aided Unicast (LC-U) }\label{sec: Unicast}

In LC-U, 
the encoder is equivalent to $n$ independent fixed-to-variable source encoders, each depending only on the local cache of the corresponding receiver.
The resulting codeword hence corresponds to $n$ unicast transmissions.
Each receiver's caching and transmission rates, $M_{i,j}$ and $R_{i,\dbf}$ $\forall (i,j) \in [n]\times[j]$, are computed as follows:

\begin{enumerate}
\item In the caching phase, each receiver computes the optimal cache allocation that minimizes the expected distortion, assuming that the sender doesn't transmit further video layers ($R_{i, \dbf} = 0$). Since users are not expecting to receive further layers during the transmission phase, each receiver caches video layers independently, based on its own demand distribution. Receiver $i \in [n]$ solves the following convex optimization problem: 
\begin{equation}
	\begin{aligned}
	&\text{min}
	& & E(D_{i}) = \sum_{j=1}^{m} q_{i,j} \sigma_{j}^{2} 2^{-2M_{i,j}} \\
	& \text{s.t}
	& & \sum_{j=1}^{m} M_{i,j} \leq M_{i}\\
	&&&  M_{i,j} \geq 0, \hspace{1cm} \; \forall j \in [m] \ 	\end{aligned}\label{eq: LCU-Cache}
\end{equation}

The resulting cache allocations are given by

\begin{equation}
M_{i,j}^{*}=  \left( \log_{2}\sqrt{\frac{2\ln{2}q_{i,j}\sigma_{j}^{2}}{\lambda_{i}^{*}}}\right)^{+}  \label{eq: LCU-CacheSol},
\end{equation}
with $\lambda_{i}^{*}$ such that $\sum_{j}M_{i,j}^{*}=M_{i}$. The solution admits a reverse water-filling form \cite{cover}. User $i$ only stores portions of video files with $q_{i,j}\sigma_{j}^{2}$ less than $\frac{\lambda_{i}^{*}}{2\ln{2}}$; hence, $q_{i,j}M_{i,j}^{*} = \min\{\frac{\lambda_{i}^{*}}{2\ln{2}}, \; q_{i,j}\sigma_{j}^{2}\}$, as illustrated in Fig. \ref{fig: WaterFilling}.

\item During the transmission phase, the optimal transmission rates for demand $\dbf \in \mathcal D$, $R_{i,\dbf}$, are derived at the sender by solving:

\begin{equation}
	\begin{aligned}
	&\text{min}
	& & \frac{1}{n} \sum_{i=1}^{n}\sigma_{d_{i}}^{2}2^{-2(M_{i,d_{i}}^{*}	   +R_{i,\dbf})} \\
	& \text{s.t.}
	& & \sum_{i=1}^{n} R_{i,\dbf} \leq R\\
	&&&  R_{i,\dbf} \geq 0, \hspace{1cm} \forall i \in [n]
	\end{aligned}\label{eq: LCU-Rate}
\end{equation}
\noindent The optimal rates are then given by 

\begin{equation}
 R_{i,\dbf}^{*}=\left( \log_{2}\sqrt{\frac{2\ln{2}\sigma_{d_{i}}^{2}}{\gamma_{\dbf}^{*}}}-M_{i,d_{i}}^{*}\right)^{+},
 \end{equation}
with $\gamma_{\dbf}^{*}$ chosen such that $\sum_{i}R_{i,\dbf}^{*}=R$.

\begin{figure}
   \centering
   \includegraphics[width=2.9in]{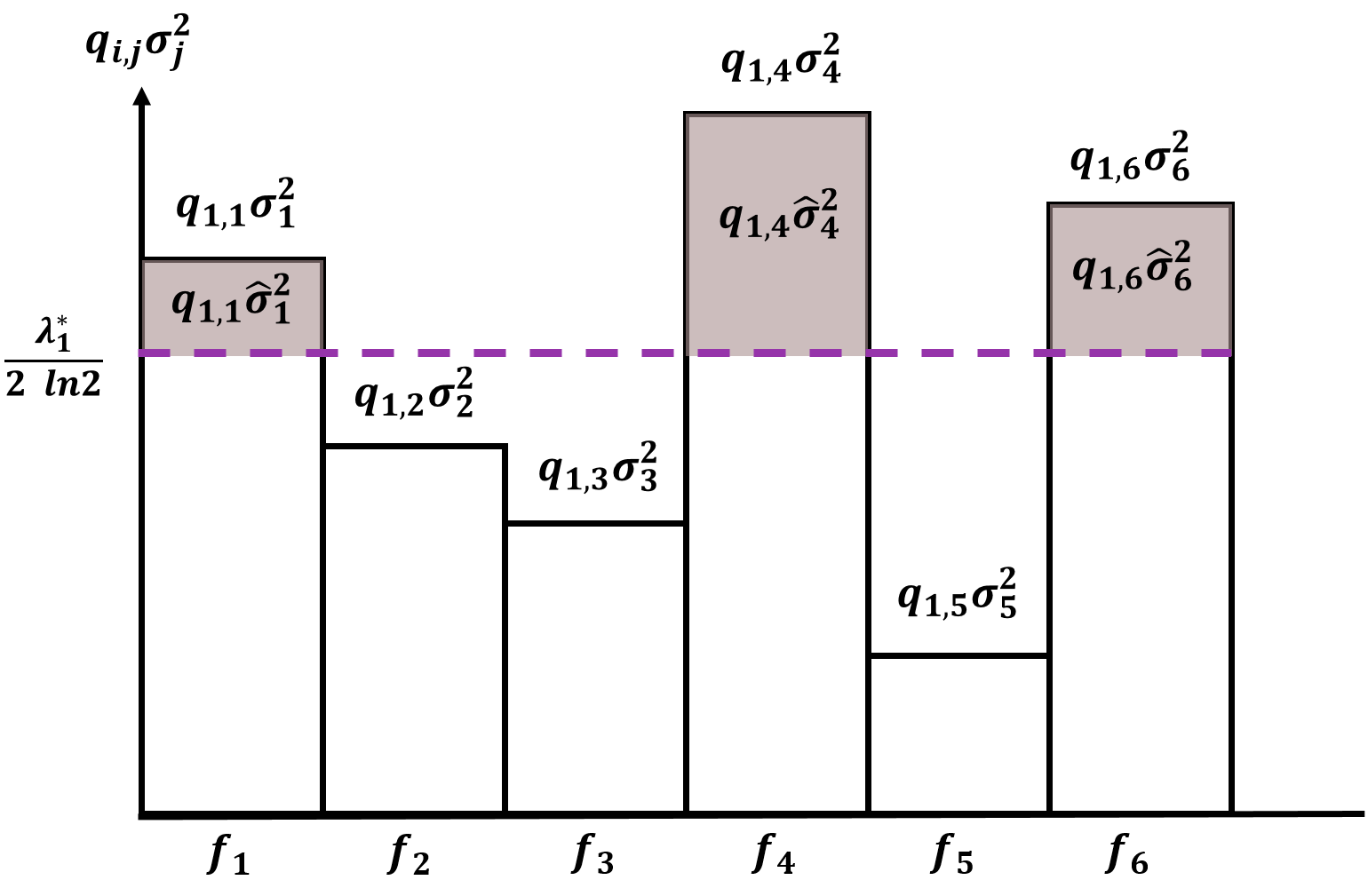}
   \caption{Illustration of the local caching scheme for $6$ files at user $1$, assuming independent Gaussian sources (video files) when $R=0$.}
   \label{fig: WaterFilling}
\end{figure}
\end{enumerate}

Note that receivers cache based on their own preferences, $q_{i,j}$, and file characteristics, $\sigma_{j}^{2}$. The caching process is decentralized and does not require any coordination from the sender.
On the other hand, delivery of the requested video files is done in a centralized manner. The sender gathers information of the cached video layers at the receivers, and jointly optimizes the transmission rates. 

\subsection{Implementation of LC-U via SVC}
LC-U is a scalable video coding scheme described by two layers, one base layer and one enhancement layer. During the caching phase receiver $i$ stores a base layer of video file $j$ at rate $M_{i,j}$ (bits/source-sample) $\forall j \in [m]$.
 In the transmission phase, the sender unicasts 
the enhancement layers of the requested video files to the corresponding receivers, at rates $R_{i,\dbf}$ $\forall i \in [n]$  which
have  been jointly optimized across all receivers. Hence, there are
$n$ disjoint encoders at the sender for transmitting the enhancement layers.


\section{Cooperative Caching-aided Coded Multicast (CC-CM) }\label{sec: Multicast}

In CC-CM, differently from LC-U, the sender encodes the requested videos jointly across all receivers. Equivalently, there is one fixed-to-variable encoder at the sender, which jointly compresses the information that needs to be delivered to each receiver based on the cached content distributed among all of them.
The sender multicasts the compressed information over the shared link; therefore, the total available rate, $R$, is no longer simply divided among the receivers. This results in more efficient transmissions over the shared link, which, in turn, increases the decoded video qualities.

As mentioned in Sec. \ref{sec: Introduction}, it has been shown that the joint design of caching and coded multicasting enables multiplicative caching gains in terms of  the aggregate rate (or load) required on the shared link to deliver the desired
per-user rate; i.e.,  the aggregate load decreases linearly with the local cache size.\cite{CACHE,maddah2013decentralized,ji2014average, ji2015random,Pr:JiTulLloCai14,ji2015Efficient,ji2015Preserving,shanmugam2014}.
However, all coded multicast schemes available in literature are based on fixed-to-variable source encoding, designed to minimize the aggregate load on the shared link so that the requested files are recovered in whole and with zero distortion.
In other words, the per-user rates are given as constraints, and the goal is to minimize the required aggregate rate over the shared link.
In contrast, in this work, we constrain the overall rate that can be transmitted over the shared link, and optimize the per-user rates, which in turn determine the receivers' video playback qualities. 

The per-user rate (in bits/source-sample) delivered to receiver $ i \in [n]$, is denoted by $R_{i, \dbf}$. We split $R_{i, \dbf}$ into two portions: a portion, $\widetilde R_{i, d_{i}} $  delivered  via coded multicast, and a portion, $\widehat R_{i, \dbf}$  delivered via unicast, so that $R_{i, \dbf}=  \widetilde R_{i, d_{i}}  + \widehat R_{i, \dbf}$ $\forall i \in [n]$.
As described later,  the  per-user coded multicast rate $\widetilde R_{i, d_{i}} $  is computed independently from the entire demand vector realization  $\dbf \in \mathcal D$.
In the following, we use \emph{aggregate coded multicast rate} to refer to the total load over the shared link associated to the multicast transmission used to deliver the per-user coded multicast rates $\widetilde R_{i, d_{i}}$, $\forall i \in [n]$,
and \emph{aggregate unicast rate} to refer to the total load over the shared link associated to the unicast transmission used to deliver the per-user unicast rates $\widehat R_{i, \dbf}$, $\forall i \in [n]$.
Note that while the aggregate unicast rate is equal to $ \sum_{i=1}^n \widehat R_{i, \dbf}$,
the aggregate coded multicast rate is, in general, much smaller than $ \sum_{i=1}^n \widetilde R_{i, d_{i}}$ (due to multicasting gains), and depends on the specific scheme adopted for the multicast transmission.
The caching-aided coded multicast scheme used in CC-CM is described in the following. 


\subsection{Random Popularity-based Caching with Greedy Index Coding}
\label{sec:acheivable}

The scheme adopted in CC-CM for  the multicast transmission is the achievable scheme proposed in \cite{ji2015random},  referred to as RAndom Popularity-based (RAP) caching with 
Greedy Constrained Coloring (GCC), or RAP-GCC.

\begin{enumerate}
\item During the caching phase, the sender computes the optimal  per-user caching and coded multicast transmission rates, $M_{i,j}$ and $R_{i,j}$ $\forall (i,j) \in [n]\times[j]$, as described in Sec. \ref{sec:RAP-GCCOptimization} as a function of the link capacity $R$,  the cache size  $M_i$  $\forall i \in [n]$, and the demand distribution $\Qbf = [q_{i,j}]$  $\forall (i,j) \in [n]\times[j]$,
using   (\ref{eq: ObjectiveGeneral})-(\ref{eq: VariableGeneral}).
Next, receiver $i$ caches $FM_{i,j}$ bits uniformly at random from the first
$F(M_{i,j}+\widetilde R_{i,j})$ bits of  file $j\in [m]$.

\item
During the transmission phase, the sender delivers the remaining $F\widetilde R_{i,d_i}$ bits, with $i\in [n]$, via coded multicast transmissions using the GCC scheme \cite{ji2015random}. Specifically,  given a demand realization ${\dbf}$, the sender
computes the multicast codeword as a function of receiver's caches and demand realization ${\dbf}$ using a (centralized) index coding based transmission scheme implemented via  GCC.  Next, the sender uses its remaining available rate to unicast further layers with rate $\widehat R_{i,{\bf d}} $ with $i\in [n]$.

\end{enumerate}

Let $R_{\dbf}^{\rm GCC}$ denote the  aggregate coded multicast rate when using RAP-GCC.
In the following, we first provide  a closed-form expression for  $R_{\dbf}^{\rm GCC}$ as a function of $M_{ij}$  and ${\widetilde R_{i,j}}$ (see Sec, \ref{sec:Achievable Rate}). This expression is then used in the optimization problem described in Sec. \ref{sec:RAP-GCCOptimization} to evaluate the optimal  $M_{ij}$,  ${\widetilde R_{i,j}}$ and  $\widehat R_{i, \dbf}$ that minimize the total average distortion.  A detailed description of the implementation of CC-CM in the context of SVC is given in Sec. \ref{implementation}.


\subsection{Achievable Rate }
\label{sec:Achievable Rate}
In the next two theorems we provide closed form expressions for the asymptotic rate (as $F$ goes to infinity) of the  RAP-GCC scheme. 
Specifically, Theorem \ref{thm:up}, quantifies the rate achieved with RAP-GCC for a demand realization $\dbf$, $R_{\dbf}^{\rm GCC}$, while Theorem \ref{thm:down} gives the average rate over all demand realizations, $R^{\rm GCC}$. These rate expressions are, then, used in the optimization problems given in Sec. \ref{sec:RAP-GCCOptimization} for finding the optimal (in the sense of minimum total average distortion) values of $M_{ij}$,  $\widetilde R_{i,j} $ and  $\widehat R_{i, \dbf}$.
For convenience, in the following, we let
\begin{eqnarray}
p^c_{i,j} =
 p_{i,j}\frac{M_{i}}{M_{i,j}+{\widetilde R_{i,j}}}.
\end{eqnarray}

\begin{theorem}
\label{thm:up}
For a shared link network with $n$ receivers, library size $m$, and cache capacity $M_i \; \forall i\in[n]$, fix a caching distribution $\pbf_i$ (or equivalently a caching placement $\Mbf_i$) $\forall i\in[n]$ and a demand realization $\dbf$. Then, the  aggregate coded multicast rate (in bits/source-sample) required to deliver $\widetilde R_{i,d_{i}} $ (bits/source-sample) to each receiver $i \in [n]$,  as $F \rightarrow \infty$, is  given by:
\begin{equation}
\label{eq:demanrate}
R_{\dbf}^{GCC} \left( \left\{ \pbf_i  \right\}_{i=1}^n \right)   \triangleq  \min \left \{   \psi_{\dbf} \left (\pbf \right),  \bar m_\dbf  \right\},
\end{equation}
\noindent where
\begin{equation}
\label{eq:demanpsi}
\psi_{\dbf} \left (\pbf \right)=\sum\limits_{\ell =1}^{n}  \sum\limits_{\mathcal{U}_{\ell}\subseteq\mathcal{U}} \max\limits_{i\in\mathcal{U}_{\ell}} \{\lambda(i,d_i)(M_{i, d_{i}}+  \widetilde R_{i, d_{i}})\},
\end{equation}
\noindent
and where $\bar m_\dbf $
is the rate (in bits/source-sample) sent via uncoded (naive) multicast given by
\begin{equation}
\label{eq:md}
\bar m_\dbf=\sum_{f=1}^m  1 \left \{  f \ni  \dbf   \right \} \max_i \left (M_{i, f}+  \widetilde R_{i, f} \right ),
\end{equation}
and finally
\begin{equation}
\label{eq:lambda}
\lambda(i,d_i) = (1-p_{i,d_{i}}^{c})\prod\limits_{u\in \mathcal{U}_{\ell}\backslash \{i\}}(p_{u,d_{i}}^{c})\prod\limits_{u\in \mathcal{U} \setminus  \mathcal{U}_{\ell}}{(1-p_{u,d_{i}}^{c})}
\end{equation}
denotes the probability that a packet from file $d_{i}$ requested by receiver $i$ has been cached by exactly $\ell-1$ receivers, 
where $\mathcal{U} = \left\{1, \ldots, n \right\}$ is the set of all receivers and $\mathcal{U}_{\ell}$ denotes a given set of $\ell$ receivers.
\hfill  $\square$
\end{theorem}

Averaging over all possible demand realizations we obtain the following result:
\begin{theorem}
\label{thm:down}
For the shared link network with $n$ receivers, library size $m$, cache capacity $M_i$ and demand distribution $\qbf_i$ $\forall i\in[n]$, fix a caching distribution $\pbf_i$ $\forall i\in[n]$.
Then, for all $\epsilon > 0$, the average aggregate coded multicast rate  (in bits/source-sample) required to deliver $\widetilde R_{i,j} $ (bits/source-sample) to each receiver $i \in [n]$ requesting file $j$,  as $F \rightarrow \infty$, satisfies:
\begin{equation}
\label{eq:2}
\lim_{F \rightarrow \infty} \Pm \left ( R^{\rm GCC} \leq
\min \left \{ \psi \left (\{\qbf_i, \pbf_i\}_{i=1}^n \right), \bar{m}  \right\} + \epsilon \right )  = 1,
\end{equation}
where
\begin{equation}
\label{eq:psi} \small
\psi \left ( \displaystyle \{\qbf_i, \pbf_i\}_{i=1}^n \right)\eqdef
\sum_{\ell=1}^n \sum_{\Uc^\ell \subset \Uc}  \sum_{f=1}^m \sum_{u \in \Uc^\ell}
\gamma_{f, u, \Uc^\ell}  \, \lambda(u,f)(M_{u, f}+  \widetilde R_{u, f}),
\end{equation}
\noindent and
\begin{equation}
\label{eq:mbar}
\bar m=\sum_{f=1}^m \left(1 - \prod_{i=1}^n \left(1 - q_{i,f}\right) \right) \max_i (M_{i, f}+  \widetilde R_{i, f}),
\end{equation}
\noindent with $\lambda(u,f) $ given as in \eqref{eq:lambda}, and where
\begin{eqnarray}\label{eq:gamma}
\gamma_{f, u, \Uc^\ell} \eqdef
\mathbb \Pm(f = \argmax\limits_{f_u \in \fbf(\Uc^\ell)} \,\,\, \lambda(u,f_u)(M_{u, f_u}+  \widetilde R_{u, f_u}))
\end{eqnarray}
denotes the probability that $f$ is the file 
that maximizes the term $\lambda(u,f)$ among $\fbf(\Uc^\ell)$, which is the set of files requested by receivers $\Uc^\ell$.
\hfill  $\square$
\end{theorem}

We remark that, the proofs for Theorems \ref{thm:up} and \ref{thm:down} omitted in this paper due to space limitations, are based on an extension of the procedure in Appendix A in \cite{ji2015random} for heterogeneous cache sizes, popularity distributions, and file sizes.

\subsection{CC-CM caching and transmission rate optimization}\label{sec:RAP-GCCOptimization}

Given the achievable scheme from Sec. \ref{sec:acheivable}, the objective is to design the caching distributions $\pbf_i$, the set of per-user coded multicast rates $\widetilde R_{i,j}$, and the set of per-user unicast  rates $ \widehat R_{i, \dbf}$ $\forall (i,j,\dbf) \in [n] \times [m] \times \mathcal D$, such that the average distortion is minimized and the channel capacity constraint is not violated.


Using $R_{\dbf}^{GCC}$ given in Eq. (\ref{eq:demanrate})-(\ref{eq:lambda}) of Theorem \ref{thm:up}, the  optimal per-user caching,  coded multicast and unicast rates,  $M_{i,j}, \widetilde R_{i, j}, \widehat R_{i, \dbf}$ are derived via the following optimization:
\begin{subequations}\label{eq: MostGeneral}
\begin{alignat}{4}
& \text{min}
&& \sum_{\dbf \in \mathcal D} \Pi_{\dbf} \bigg(\frac{1}{n} \sum_{i=1}^{n}\sigma_{d_{i}}^{2}2^{-2(M_{i, d_{i}}+\widetilde R_{i, d_{i}}+ \widehat R_{i, \dbf})}\bigg)\label{eq: ObjectiveGeneral}\\
& \text{s.t.}
&& \min \left \{   \psi_{\dbf} \left (\pbf \right),  \bar m_\dbf  \right\} + \sum\limits_{i=1}^{n}  \widehat R_{i, \dbf}\leq R,   \hspace{0.5cm} \forall \dbf \in \mathcal D \label{eq: RateGeneral}\\
&&& \sum_{i=1}^{n}M_{i,j} \leq M_{i}, \hspace{1cm} \forall i \in [n]\label{eq: CacheGeneral}\\
&&& M_{i,j}, \widetilde R_{i, j}, \widehat R_{i, \dbf}\geq 0,  \hspace{0.3cm}  \forall (i, j, \dbf) \in [n]  \times [m] \times \mathcal D \label{eq: VariableGeneral}
\end{alignat}
\end{subequations}

Eq. \eqref{eq: RateGeneral} corresponds to the rate constraint, with its first term being 
the aggregate average coded multicast rate achieved with RAP-GCC. 
The second term in \eqref{eq: RateGeneral} is the aggregate  unicast rate for demand $\dbf$.

The optimization problem in \eqref{eq: MostGeneral} is highly non-convex and has an exponential number of constraints due to \eqref{eq: RateGeneral}, which depends on the cardinality of $\mathcal D$. In order to reduce complexity, we allow satisfying the rate constraint on average over all demands rather than for each demand realization, and replace \eqref{eq: RateGeneral} with the following expression:
\begin{equation}\label{subeq:AverageRate}
\min \left \{ \psi \left (\{\qbf_i, \pbf_i\}_{i=1}^n \right), \bar{m}  \right\} + \sum_{\dbf \in \mathcal D} \Pi_{\dbf}\sum\limits_{i=1}^{n} \widehat R_{i, \dbf}\leq R,
\end{equation}
where $\psi \left (\{\qbf_i, \pbf_i\}_{i=1}^n \right)$ and $ \bar{m} $ are given by Eq.(\ref{eq:psi})-(\ref{eq:gamma}) in Theorem \ref{thm:down}.

\subsection{Implementation of CC-CM via SVC}
\label{implementation}

In order to implement the CC-CM scheme via SVC, the following steps are executed:
\begin{enumerate}
\item The optimal caching and transmission rates, $M_{i,j}$, $\widetilde R_{i, j}$, $\widehat R_{i, \dbf}$, are computed by solving the optimization problem  (\ref{eq: ObjectiveGeneral})-(\ref{eq: VariableGeneral}) with   \eqref{eq: RateGeneral}  replaced by \eqref{subeq:AverageRate}.
\item The  videos are split into multiple layers and each layer is encoded at the same rate, $b$ bits/source-sample, represented as a binary vector of length (entropy) $bF$ bits.  The value of $b$ is chosen as follows: Let
$$\mu_{i,j} \eqdef \frac{M_{i,j}}{b}, \qquad
\widetilde \rho_{i, j} \eqdef \frac{\widetilde R_{i, j} }{b}, \qquad
 \widehat \rho_{i, \dbf}  \eqdef  \frac{\widehat R_{i, \dbf}}{b},
$$
then, the common layer rate b (bits/source-sample) is  such that   $\mu_{i,j} $, $\widetilde \rho_{i, j}$, and $\widehat \rho_{i, \dbf}$ are integers  $\forall (i, j, \dbf) \in [n]  \times [m] \times \mathcal D$.
Note that $\mu_{i,j}$ denotes the number of layers of video file $j$ cached by receiver $i$, $\widetilde \rho_{i, j} $ represents the number of layers delivered via coded multicast,  while  $ \widehat \rho_{i, \dbf}$ denotes the number of layers that are unicast to receiver $i$. Finally, the sum $\omega_{i,j} = \mu_{i, j}  + \widetilde \rho_{i, j}$,
referred to as  the  \emph{storing range} of receiver $i$ for video file $j$, indicates the highest
layer\footnote{
We index video layers according to their sequential order in SVC,
where layer $k$ is only useful if layers $\{1,\ldots,k-1\}$ are present.}
of video file $j$ that receiver $i$ is allowed to cache.
In other words, $\omega_{i,j}b$ represents  the {\em average} rate  (in bits/source-sample) guaranteed to receiver $i$ after the coded multicast transmission.



\item
Each receiver configures its cache based on the implementation of RAP  via SVC described in Sec. \ref{rap}.I.
\end{enumerate}

The above steps are all part of the caching phase.
During the delivery phase, at each use of the network, after a demand realization $\bf d$ is generated, the sender uses the implementation of GCC  via SVC described in Sec. \ref{gcc}.II.

\subsection*{ \mbox{\rm D.I)} RAP via SVC:}~
\label{rap}
As in \cite{CACHE,ji2014average,ji2015random}, in order to maximize coded multicast  opportunities, file layers are divided into $B$ equal size packets each with $b F/B$  bits. Receiver $i \in [n]$ caches $\mu_{ij}B$ packets uniformly at random from the $\omega_{i,j}B$ packets of the first $\omega_{i,j}$ layers of video file $j \in [m]$. In this context, $p^c_{i,j}$ admits the interpretation of the probability that a packet from the first $\omega_{i,j}$ layers of video file $j$ is  cached at user $i$, i.e.:
\begin{eqnarray}
p^c_{i,j} \eqdef p_{i,j}\frac{M_{i}}{M_{i,j}+{\widetilde R_{i,j}}} =\frac{\mu_{i,j}}{\omega_{i,j}} =\frac{\binom{\omega_{i,j}B-1}{\mu_{i,j}B-1}}{\binom{\omega_{i,j}B}{\mu_{i,j}B}}.
\end{eqnarray}

Fig. \ref{fig: LayerPacket} illustrates the layer-packet division of a video file stored at a receiver. The layers are encoded at the same rate $b$ bits/sample and each layer of length $bF$ bits is divided into 5 packets. The number of cached layers is $\mu = 3$ and the storing range for this video file is $\omega = 6$. The receiver has cached 15 packets from the 30 packets forming the first 6 layers.

{
We remark that, in contrast to LC-U, in which the cached content can be configured locally by each receiver, the caching scheme in CC-CM requires coordination from the sender, which computes the caching rates jointly across all receivers, as described in Sec. \ref{sec:RAP-GCCOptimization}. 
}

\begin{figure}
   \centering
   \includegraphics[width=3.3in]{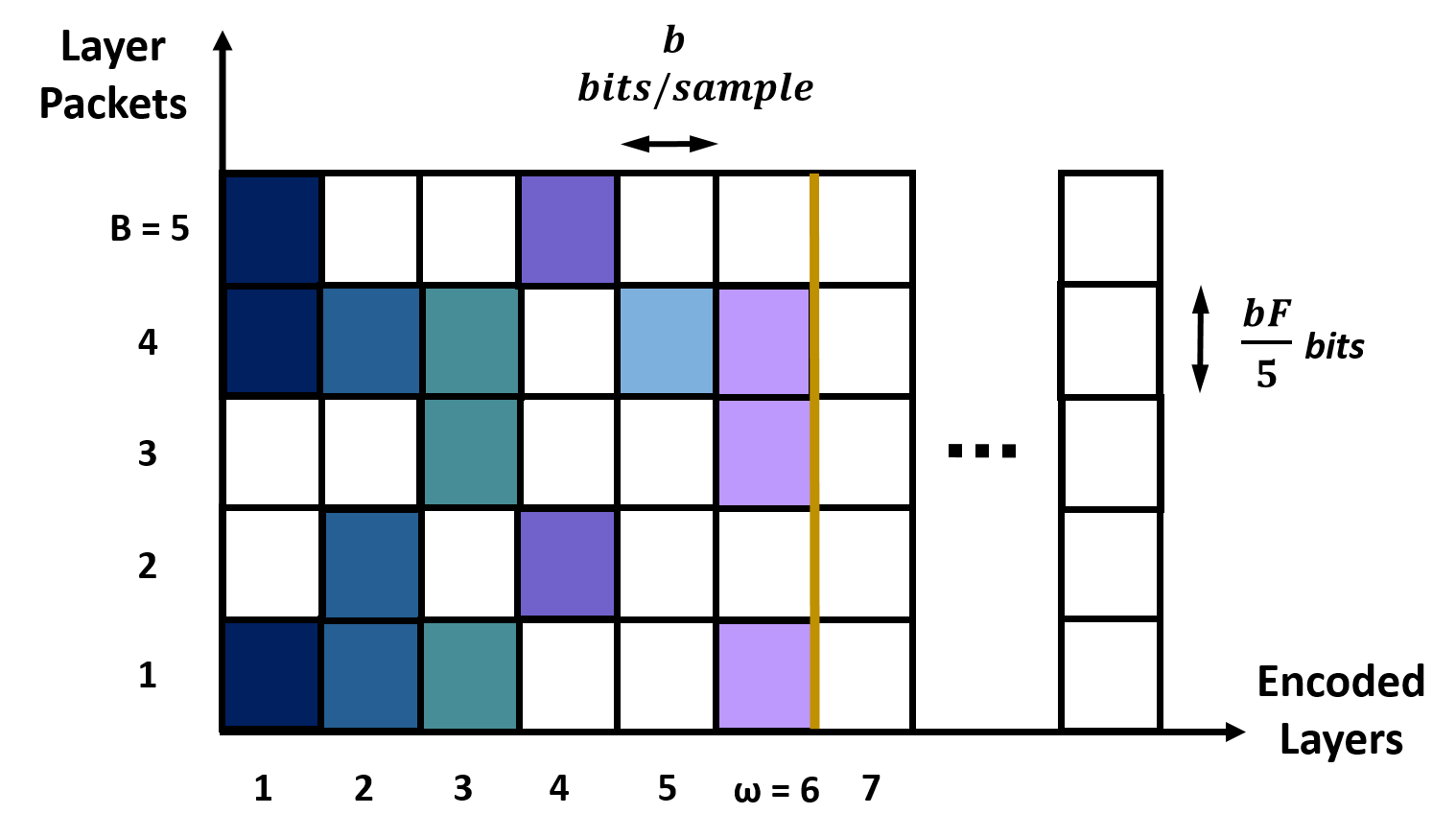}
   \caption{Layer-packet division of a file where $\mu = 3$ and $\omega = 6$. Files are encoded at multiple layers (represented with different colors). Each layer has rate $b$ (bits/sample) and  consists of B = 5 packets. The receiver caches 15 packets from a total of 30 packets of the first $6$ layers. }
   \label{fig: LayerPacket}
\end{figure}

\subsection*{ \mbox{\rm D.II) } GCC  plus Unicast Transmission via SVC: }~
\label{gcc}
For a given cache placement and demand realization $\dbf$,
the sender identifies the set of packets 
that need to be delivered via coded multicast so that each receiver $i\in[n]$ is guaranteed  {\em  average} rate  $\omega_{i,d_i}b$. Specifically,  the sender has to deliver  $ \widetilde \rho_{i, d_i}B = (\omega_{i,d_i}  - \mu_{i,d_i})B$ packets, or equivalently, $\widetilde R_{i, d_i}= \widetilde \rho_{i, d_i}b$ bits/sample.
In CC-CM, the sender uses the GCC coded multicast 
transmission scheme, which is  based on a greedy solution for the minimum vertex coloring of the corresponding index coding conflict graph \cite{bar2011index},
whose detailed description can be found in \cite{ji2014average, ji2015random}.
Having exploited all coded multicast opportunities for demand $\bf d$, receivers will have received the first $\omega_{i,d_{i}}$ layers of their requested video files with rate $\widetilde R_{i,d_{i}} $. The sender, then, uses its remaining available rate to unicast further layers with rate $\widehat R_{i,{\bf b}} $.

It is important to remark that in SVC, a layer is used as part of the decoding process only if all its preceding layers are available.
Therefore, one needs to ensure that after completion of the transmission phase, receiver $i \in [n]$ has up to the ($\mu_{i,d_{i}}+\widetilde \rho_{i, d_i}+ \widehat \rho_{i, \dbf})-th$ layer of file $d_i$.
However, since
for reducing complexity, the coded multicast rates, $\widetilde R_{i, d_i}=\widetilde \rho_{i, d_i}b$, are computed using  the average rate constraint \eqref{subeq:AverageRate}, there may be specific  demand realizations for which the  link capacity $R$ is violated. Consequently the sender is not able to deliver all the  $ \widetilde \rho_{i, d_i}B$ packets that ensure reconstruction of the $ \omega_{i, d_i}$ layers by receiver $i$.
In such cases, CC-CM  employs a greedy approach, in which the per-user coded multicast rates, $\widetilde R_{i, d_i}$, are reduced in decreasing order of  $\sigma_j^2$ until the link capacity constraint is satisfied.

\section{CC-CM Optimization: Simplifications }\label{sec:RAP-GCCsimplified}
In this section, we focus on optimization problem (\ref{eq: MostGeneral}) with  rate constraint (\ref{eq: RateGeneral}) replaced by (\ref{subeq:AverageRate}), and assume key network symmetries which allows us to quantify the solutions.


\subsection{Symmetry Across Users}~\label{sec:Symmetrical}
We assume that receivers have the same cache size and they request video files according to the same distribution, i.e $M_i=M$ and  $q_{i,j}=q_{j}$, $\forall (i , j) \in [n]\times [m]$. Hence, without loss of optimality, we can assume that the caching distributions $\pbf_i$, and the corresponding caching placements $\Mbf_i$ are constant across all receivers (i.e., $\Mbf_i=\Mbf$). Analogously, we can assume that all receivers cache $\mu_jB$ packets of video file $j$ up to the same storing range $\omega_{j}$, or equivalently, the optimal rate delivered to each receiver via coded multicast for $d_i=j$ is $\widetilde R_{i,j}=\widetilde R_{j}$ $\forall i \in [n]$.
Therefore, the probability of a packet from file $j$ being cached at any receiver is $p_{j}^{c} = \frac{\mu_{j}}{\omega_{j}}= \frac{M_{j}}{M_{j}+\widetilde R_{j}}$. The average aggregate rate required for delivering the per-user coded multicast rate $\widetilde R_{j}$ (bits/source-sample) to any receiver requesting file $j$, given by \eqref{eq:2}, is  simplified as in \cite{ji2015random} to:
\begin{eqnarray}
R^{\rm GCC} =
\min \left \{ \psi \left (\qbf, \pbf \right), \bar{m}  \right\},
\end{eqnarray}
\noindent where
\begin{equation}
\bar m \eqdef \sum_{j = 1}^m \left(1 - \left(1 - q_j \right)^{n} \right)(M_{j}+\widetilde R_{j}),
\end{equation}
\begin{equation}
\psi \left (\qbf, \pbf \right)=\sum\limits_{l =1}^{n} \binom{n}{l} \sum\limits_{j=1}^{m} \gamma_{j,\ell}(p_{j}^{c})^{l-1}(1-p_{j}^{c})^{n-l+1}(M_{j}+\widetilde R_{j}),
\end{equation}
\vspace{-0.2cm}
and
\begin{equation}
\gamma_{j,\ell} \triangleq \mathbbm{P} \bigg( j = \argmax\limits_{f\in\mathcal{D}} (p_{f}^{c})^{l-1}(1-p_{f}^{c})^{n-l+1}(M_{f}+\widetilde R_{f})\bigg),
\end{equation}
with $\mathcal{D}$ being the random subset of $l$ files chosen in an i.i.d manner from the library (with replacement).

We refer to the scheme resulting from the solution of the following optimization problem as CC-CM using RAP-GCC: 
\begin{subequations}\label{eq: Symplified}
\begin{alignat}{4}
& {\text{min}}
\hspace{0.2cm} && \sum_{\dbf \in \mathcal D} \Pi_{\dbf} \bigg(\frac{1}{n} \sum_{i=1}^{n}\sigma_{d_i}^{2}2^{-2(M_{d_i}+\widetilde R_{d_i} + \widehat R_{i, \dbf})} \bigg) \\
& \text{s.t.}
&& \min \left \{ \psi \left (\qbf, \pbf \right), \bar{m}  \right\} + \sum_{\dbf \in \mathcal D} \Pi_{\dbf}\sum\limits_{i=1}^{n} \widehat R_{i, \dbf} \leq R, \\ \label{subeq:RateSimplified}
&&& \sum\limits_{j=1}^{m}M_{j} \leq M \\
&&& M_{j}, \widetilde R_{j}, \widehat R_{i, \dbf}\geq 0,  \hspace{0.3cm}  \forall (i, j, \dbf) \in [n]  \times [m] \times \mathcal D
\end{alignat}
\end{subequations}

As proposed in \cite{ji2015random}, the caching placement can be simplified according to the following truncated uniform caching distribution:
\begin{equation}
p_j = \begin{cases} \frac{1}{\widetilde m},  &\hspace{0.5cm}  j \leq \widetilde m \\ 0, \hspace{2cm}&\hspace{0.5cm}  j \geq \widetilde m + 1 \end{cases},
\end{equation}
where the cut-off index $\widetilde m \geq M$ is a function of system parameters.
The resulting caching scheme is referred to as the Random LFU (RLFU) caching scheme \cite{ji2015random}. RLFU caching
is equivalent to all receivers uniformly caching the most $\widetilde{m}$ popular videos files. Therefore:
\vspace{-0.7cm}
\begin{multicols}{2}
    \begin{equation}\nonumber
        M_{j} = \begin{cases} \widetilde M = \frac{M}{\widetilde m } &\hspace{0.1cm}  j \leq \widetilde m \\ 0 \hspace{1cm}&\hspace{0.2cm}  j \geq \widetilde m + 1 \end{cases},
    \end{equation}\break
    \vspace{-0.5cm}
    \begin{equation}\nonumber
    \widetilde R_{j}  = \begin{cases} \widetilde R   &\hspace{0.2cm}  j \leq \widetilde m \\ 0 \hspace{1cm}&\hspace{0.2cm}  j \geq \widetilde m + 1 \end{cases}.
    \end{equation}
\end{multicols}
\noindent A packet of file $j$ is cached at any receiver with probability:
\begin{equation}
p^c_j = \begin{cases} \frac{\widetilde M}{\widetilde M+\widetilde R}  &\hspace{0.5cm}  j \leq \widetilde m \\ 0 \hspace{2cm}&\hspace{0.5cm}  j \geq \widetilde m + 1 \end{cases},
\end{equation}
\noindent which results in the following optimization problem:
\begin{subequations}\label{eq: RLFU}
\begin{alignat}{4}
&{\text{min}}
\hspace{0.2cm} && \sum_{\dbf \in \mathcal D} \Pi_{\dbf} \bigg(\frac{1}{n} \sum_{i=1}^{n}\sigma_{d_i}^{2}2^{-2(M_{d_i}+\widetilde R_{d_i} + \widehat R_{i, \dbf})} \bigg) \\
& \text{s.t.}
&& \left (\frac{\widetilde R }{\widetilde M + \widetilde R} \right) \left (1-\left (\frac{\widetilde R}{\widetilde M + \widetilde R} \right)^{nG_{\widetilde m}} \right )(\widetilde M + \widetilde R) + \nonumber \\
&&& n (1-G_{\widetilde{m}})\widetilde R +\sum_{\dbf \in \mathcal D} \Pi_{\dbf}\sum\limits_{i=1}^{n} \widehat R_{i, \dbf} \leq
R,\label{subeq:RateRLFU}\\
&&& \widetilde M, \widetilde R, \widehat R_{i, \dbf}\geq 0,  \hspace{0.3cm}  \forall (i, \dbf) \in [n]  \times \mathcal D
\end{alignat}
\end{subequations}
where $G_{\widetilde{m}} =  \sum_{j=1}^{\widetilde{m}}q_{j}$, and
the coded multicast rate expression in Eq. (\ref{subeq:RateRLFU}), is derived in Eq. (17) of \cite{ji2015random}.
We refer to the resulting scheme as  CC-CM  using RLUF-GCC. 


\subsection{Symmetry Across Users and Files (Sources)}~\label{sec: Uniform}
Finally, the simplest network setting would be for all users to have equal-size caches, to request video files uniformly and for all sources to have the same distribution, i.e.\ $M_{i} = M,\; q_{i,j} = 1/m, \;\sigma_{j}^{2} = \sigma^{2} ,\; \forall (i,j) \in [n]\times[m]$. Consequently, with no loss of optimality, we can assume that  $\widetilde R_{i,j}=\widetilde R$ and $\widetilde M_{i,j}=\widetilde M$  $\forall (i,j) \in [n]\times[m]$ which results in $p^{c} = \frac{\widetilde M}{\widetilde M+\widetilde R}$. It is immediate to see that in this setting the optimal solution assigns $\widehat R_{i, \dbf}=0$ $ (i,\dbf) \in [n]\times\mathcal D$, and only the coded multicast rates need to be accounted for.
Optimal values of $\widetilde R$ and $\widetilde M$  in terms of minimum average distortion are given by
\begin{equation}
	\begin{aligned}
&{\text{min}}
	& & \sigma^{2}2^{-2(\widetilde M+\widetilde R)} \\
	& \text{s.t.}
	& &   \frac{\widetilde R}{\widetilde M} \left (1- \left (\frac{\widetilde R}{\widetilde M+\widetilde R}  \right)^n \right)(\widetilde M+\widetilde R) \leq R\\
	&&& \widetilde M \leq M\\
    &&& \widetilde M, \widetilde R \geq 0
	\end{aligned}\label{eq: Uniform}
\end{equation}

\vspace{0.2cm}
\section{Simulation Results}~\label{sec:Simulations}
In this section, we numerically compare the performance of the unicast and multicast caching-aided transmission schemes introduced in Sec. \ref{sec: Unicast} and \ref{sec: Multicast} for the simplified RAP-GCC settings analyzed in Sec.\ref{sec:RAP-GCCsimplified}.

We consider a network composed of $n = 20$ receivers and a library with $m = 100$ video files.
We assume that videos are requested according to a Zipf distribution with parameter $\alpha$:
$$q_{j} = \frac{j^{-\alpha}}{\sum_{f = 1}^{m} f_{-\alpha}}   \hspace{2cm}   j  \in[m].$$

In order to reduce the complexity of the RAP-GCC optimization, we assume $\widehat R_{i,\dbf}$ is independent of the demand and depends only on the file identity. Note that this assumption may lead to suboptimal solutions and hence the results in this section represent an upper bound on the optimal performance (in terms of distortion).

Fig. \ref{fig:Symmetry Across Receivers}, displays the expected average distortion achieved with the LC-U and CC-CM schemes using RAP-GCC. It is assumed that all receivers have the same cache size, $\alpha = 0.6$ and  $\sigma_{j}^{2}$ is uniformly distributed in  $[0.7,1.6]$. The distortions have been plotted (on a logarithmic scale) for  link capacity values of $R = 2,5,8$ bits/sample as receiver cache sizes vary from $5$ to $100$ bits/sample.
As expected, CC-CM significantly outperforms LC-U  in terms of average distortion.
This means that for a given shared link capacity constraint, $R$, CC-CM is able to deliver more video layers to the receivers, reducing their
experienced distortions, and equivalently, increasing their video playback quality. 
Specifically, for capacity $R= 2$ and cache size $M=50$, CC-CM achieves a $2.1\times$ reduction in expected distortion compared to LC-U. Observe that when the rate goes up to $R=8$, for the same cache size $M=50$, the gain of CC-CM increases to $5.4$.

\begin{figure}
   \centering
   \includegraphics[width=2.5in]{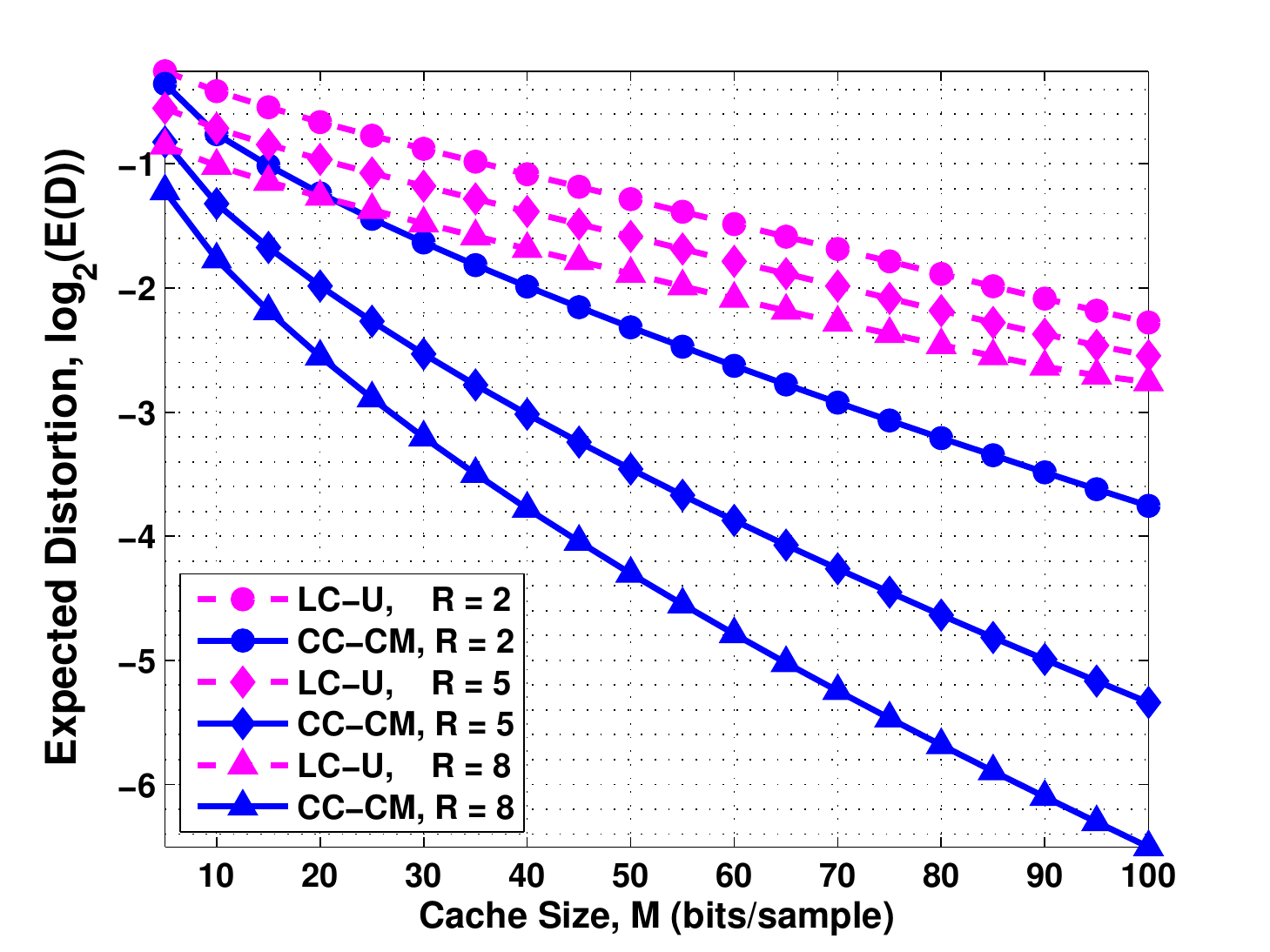}
   \caption{$n = 20$ receivers, $m = 100$ videos and Zipf  parameter $\alpha = 0.6$. For the multicast transmission scenario, caching and transmission schemes are based on RLUF and GCC respectively.}
   \label{fig:Symmetry Across Receivers}
\end{figure}

In Fig. \ref{fig:Symmetry Across Receivers and Files}, we simulate a homogeneous network scenario with $\alpha = 0$ (files are requested uniformly) and $\sigma_{j}^{2} = 1.5 $ $\forall j \in [100]$. The expected distortions achieved for LC-U and CC-CM (using RAP-GCC) are plotted for the values of $R = 2,5,10$ bits/sample as receiver cache sizes vary from $5$ to $100$ bits/sample. Observe how the gains of CC-CM are even higher in this scenario. This is due to the increased coded multicast opportunities that arise when files have uniform popularity \cite{ji2015random}. Note that in this case, for $R=10$ and $M=50$, the average distortion with CC-CM is $9.5$ times lower than with LC-U. The improvement factor goes up to $14\times$  with $M=70$.

\begin{figure}
   \centering
   \includegraphics[width=2.5in]{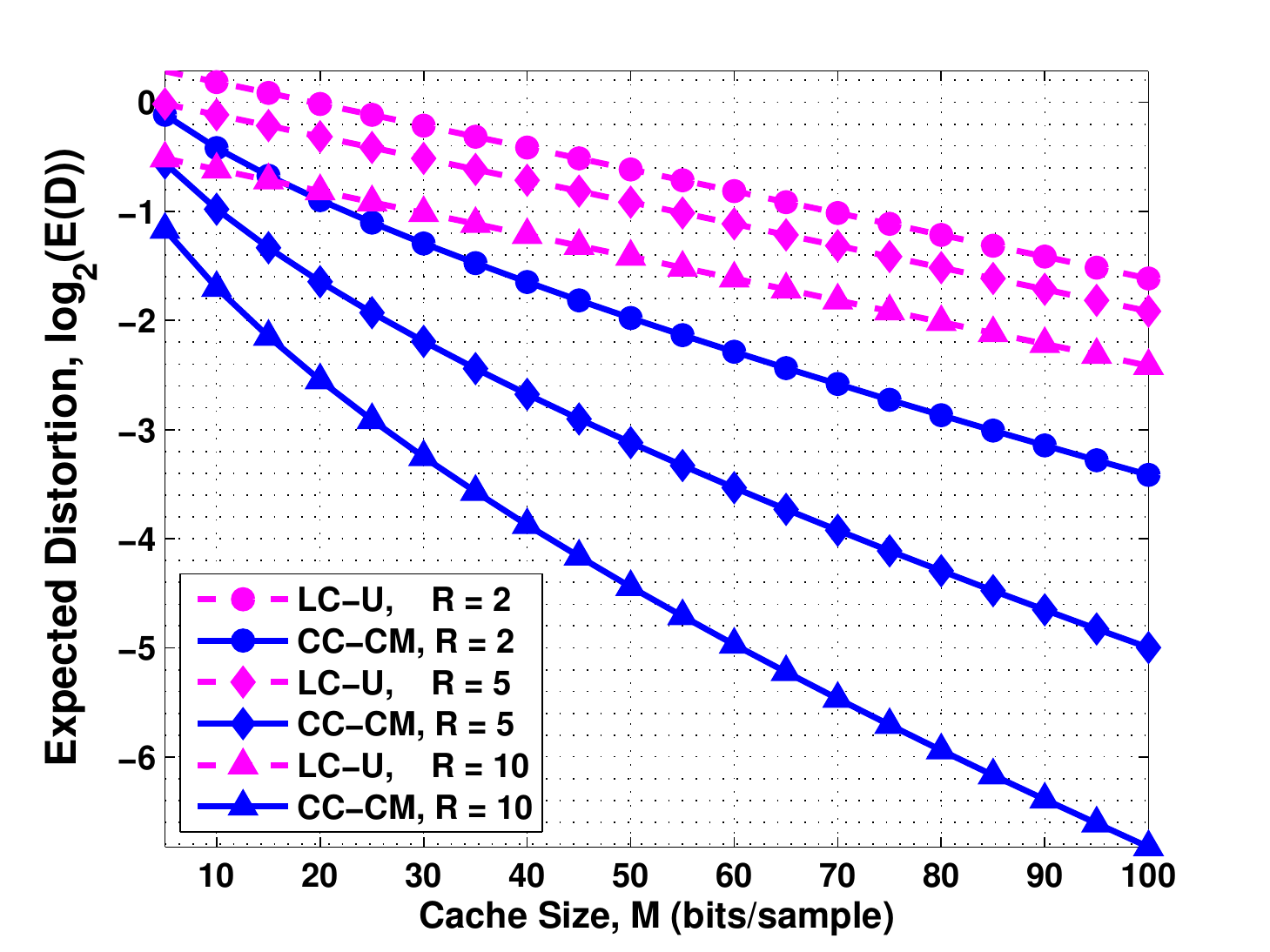}
   \caption{$n = 20$ receivers, $m = 100$ videos and  Zipf  parameter $\alpha = 0$. For the multicast transmission scenario, caching and transmission schemes are based on RAP-GCC, analyzed in Sec. \ref{sec: Uniform}.}
   \label{fig:Symmetry Across Receivers and Files}
\end{figure}

In practice, receivers cache content based on the solutions derived from the optimization problems, and the sender transmits further layers through coded multicast and naive multicast (instead of unicast) which results in distortions lower than those initially computed. We also remark that, when implementing the schemes ($B \rightarrow \infty$), the integer constraint on $\mu_{i,j}$, $\widetilde\rho_{i,j}$ and $\hat\rho_{i,\dbf}$ can be relieved to $\mu_{i,j}+\widetilde\rho_{i,j}+\hat\rho_{i,\dbf}$ being an integer $\forall (i,j,\dbf) \in [n]\times[m]\times[\mathcal D]$. For example, $\mu_{1,1}= 2.5$ means user $1$ caches the two most significant layers fully and only caches half of the packets from the third layer.


\section{Conclusion}~\label{sec: Conclusion}
In this paper, we have investigated the use of caching for enhancing video streaming quality, or in a more abstract sense, reducing source distortion. Receivers cache low rate versions of the video files and during the transmission phase further video layers are delivered to enhance the video playback quality. 
We show that while local caching and unicast transmission
results in acceptable distortion without the need of global coordination, the use of cooperative caching and coded multicast transmission is able to provide order improvements in average achievable distortion by delivering more enhancement video layers with the same available broadcast rate.


{We remark that while partitioning videos into multiple equal-size layers is key to fully exploiting multicast opportunities, the performance of SVC scheme degrades as the number of layers increases due to coding overhead. This tradeoff between multicasting gains and coding overhead is the subject of ongoing work.}



\bibliographystyle{IEEEtran}
\bibliography{QualRef,proceedings,articles,misc,references_d2d_2}

\end{document}